\documentclass[fleqn,usenatbib,useAMS]{mnras}

\usepackage{newtxtext,newtxmath}
\usepackage[utf8x]{inputenc}
\usepackage{latexsym}
\usepackage[T1]{fontenc}
\usepackage{ae,aecompl}
\usepackage{float}
\usepackage{graphicx}
\usepackage{amsmath,amsfonts}
\usepackage{lscape}
\usepackage{afterpage}
\usepackage{threeparttable}
\usepackage{mathtools}
\usepackage{pdflscape}
\usepackage[export]{adjustbox}
\usepackage[dvipsnames, table]{xcolor}

\usepackage{booktabs}

\usepackage{svg}
\usepackage{academicons}
\newcommand{\orcid}[1]{\href{https://orcid.org/#1}{\includegraphics[width=10pt]{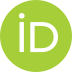}}}

\title[Millimeter-wavelength variability of YSOs]{A systematic survey of millimetre-wavelength flaring variability of Young Stellar Objects in the Orion Nebula Cluster}

\author[J. Vargas-González et al.]
{
J. Vargas-Gonz\'{a}lez \orcid{0000-0003-4329-3299},$^{1,2}$\thanks{E-mail: jvargasg@eso.org}
J. Forbrich \orcid{0000-0001-8694-4966},$^{1,3}$
V. M. Rivilla \orcid{0000-0002-2887-5859},$^4$
K. M. Menten \orcid{0000-0001-6459-0669}, $^5$
M. Güdel $^6$
and A. Hacar \orcid{0000-0001-5397-6961} $^6$
\\
$^1$Centre for Astrophysics Research, School of Physics, Astronomy and Mathematics, University of Hertfordshire, Hatfield AL10 9AB, UK\\
$^2$ESO, Alonso de Córdova 3107 Vitacura, Casilla, 19001, Santiago, Chile\\
$^3$Harvard-Smithsonian Center for Astrophysics, 60 Garden St, Cambridge MA 02138, USA\\
$^4$Centro de Astrobiología (CSIC-INTA), Ctra. de Ajalvir Km. 4, Torrejón de Ardoz, 28850 Madrid, Spain\\
$^5$Max Planck Institut für Radioastronomie, Auf dem Hügel 69, D-53121 Bonn, Germany\\
$^6$Department of Astrophysics, University of Vienna, Türkenschanzstrasse 17, A-1180 Vienna,Austria\\
}

\date{Accepted XXX. Received YYY; in original form ZZZ}

\pubyear{2022}

\begin{document}

\label{firstpage}
\pagerange{\pageref{firstpage}--\pageref{lastpage}}
\maketitle


\begin{abstract}

High-energy processes are ubiquitous even in the earliest stages of protostellar evolution. Motivated by the results of our systematic search for intense centimeter radio flares in Young Stellar Objects (YSOs) and by rare findings of strong millimeter-wavelength variability, we have conducted a systematic search for such variability in the Orion Nebula Cluster (ONC) using Atacama Large Millimeter/submillimeter Array (ALMA). Rapid variability on timescales of minutes to hours in the (centimeter)millimeter-wavelength range indicates (gyro)synchrotron radiation. Additionally, mass accretion will also affect the millimeter-wavelength luminosity but typically on longer timescales. Beyond studies of individual YSOs, our characterization of strong millimeter-wavelength variability with ALMA in the ONC sets first systematic constraints on the occurrence of such variability in a large number of YSOs ($\sim$130). We report the discovery of an order of magnitude millimeter-flare within just a few minutes from a known YSO previously reported as a radio flaring source at cm-wavelengths (the ``ORBS'' source). We also present an assessment of the systematic variability effects caused by the use of time-sliced imaging of a complex region. These are mostly due to the impact of a changing synthesized beam throughout the observations. We use simulated ALMA observations to reproduce and quantify these effects and set a lower limit for the variability that can be studied using our method in a complex region such as the ONC. Our results demonstrate that the utility of time domain analysis of YSOs extends into the millimeter-wavelength range, potentially interfering with the conversion of observed fluxes into dust masses.

\end{abstract}


\section{Introduction}\label{sec:intro}

High-energy processes are already present at the earliest stages of protostellar evolution as revealed by X-ray and radio observations (e.g., \citealt{fem99}). At radio wavelengths, these processes can be traced by nonthermal emission of (gyro)-synchrotron radiation as a result of the electron population gyrating along magnetic field lines in protostellar coronae and vicinities (innermost regions of circumstellar disks). In this context, mildly relativistic electrons can produce gyrosynchrotron radiation detectable at cm-wavelenghts, while electrons at higher energies (MeV) are responsible for synchrotron radiation into the millimeter range \citep{dul85, gue02}. Tracers of nonthermal radio emission include strong variability, negative spectral indices, and polarization. Despite their related nature, the physical connection between the emission at mm- and cm-wavelengths is just partially understood due to a lack of suitable data and, while a single source may show both simultaneously, there is evidence of millimeter-wavelength solar flares without centimeter counterparts (e.g., \citealt{kundu2000}).

Centimeter radio emission from Young Stellar Objects (YSOs) has been explored in more detail in the last few years due to the improved sensitivity of radio facilities such as the Karl G. Jansky Very Large Array (VLA) and the Very Long Baseline Array (VLBA) \citep{riv15,for16,she16,tobin2016,forbrich2021,vargas2021}. Recent analysis of deep VLA observations at cm-wavelengths towards hundreds of YSOs in the Orion Nebula Cluster (ONC) revealed intense radio flares with changes in flux density by a factor of 10 in less than 30~min and denominated as extreme radio variability events (\citealp{for17}, see also \citealp{vargas2021}). 
These studies comprise a systematic search for YSOs variability at cm-wavelengths totaling up to $\sim$7440~h of cumulative YSO observing time and leading to a mean time between extreme radio variability events of $2482\pm1433$~h. On the other hand, millimeter continuum observations of YSOs are typically used to study the thermal component of circumstellar disks that arises from dust emission assumed to be constant on short timescales. However, a few serendipitous discoveries have shown evidence of strong millimeter flares in YSOs. The first such discovery was a mm-wavelength flare towards a T Tauri star in the ONC (GMR A) as reported in \citet{bow03}. During these observations using the BIMA array at 86~GHz ($\sim$3~mm) this source became the brightest one in the cluster. This flare was coincidentally complemented with simultaneous X-ray Chandra observations that found strong X-ray activity, starting two days prior to the 3-mm flare.

An additional example of a mm-flare, found towards the T Tauri binary system V773 Tau A, was interpreted to arise from interbinary collisions of coronal structures (``helmet streamers'' of one component with the corona of the other) which results in regular flaring activity \citep{massi2002, massi2006, massi2008}. A similar interpretation has been proposed for recurring millimeter-wavelength flares in the T Tauri spectroscopic binary system DQ Tau, after the discovery of a strong flare at 3~mm that peaked at almost $\sim$0.5~Jy. Follow-up observations suggest that these flares come from synchrotron emission due to interacting protostellar magnetospheres near periastron passage \citep{salter2008, salter2010}. 

At shorter wavelengths (450 and 850~$\mu$m), a submillimeter flare was reported in \citet{mairs2019} towards the binary T Tauri system JW~566, also in Orion. It was even more luminous than the flares detected in GMR A and DQ Tau, and represents the first coronal YSO flare detected at submillimeter wavelengths. Together with the few examples of short-timescale mm flares, there are also millimeter variability studies of YSOs on longer timescales and in a different context where thermal dust emission is more relevant and its variability is caused by active mass accretion periods with an impact on timescales of months to years (\citealt{liu2018, francis2019} and references therein). 

Early estimates for the expected number of radio flares with changes in flux density greater than a factor of 5 in a few hours that can be detected in the Orion nebula at millimeter wavelengths using ALMA were as high as $\sim$10-100 flares in short integration times (minutes) for a sensitivity of $\sim$0.1~mJy and even $\sim$100-1000 flares for observations with sensitivity of $\sim$10~$\mu$Jy \citep{bow03}. Similarly, in a more specific frequency range, it has been proposed that with the high sensitivity that ALMA band 3 observations could achieve within just a few hours (on the order of $\sim$10~$\mu$Jy) in a small area in the core of the ONC ($<$30~arcsec) it would be possible to find $\sim$6 radio flares per day with change in flux density by a factor $>$2 on timescales of hours to days \citep{riv15}.
However, such sensitivity was not achieved in the ALMA band 3 observations that we are presenting here, which ranges between $\sim$100 and $\sim$300~$\mu$Jy  (see section \ref{sec:obs}). Two important elements in the search for flares in such observations are the sensitivity provided by ALMA and the large number of sources in the ONC.

Given the lack of a statistical sample of strong and short-lived millimeter flares we started a first systematic search for such events in YSOs using ALMA, targeting the BN/KL region close to the core of the ONC for a large sample of sources, and observing on short timescales of minutes to days. The use of ALMA is a major benefit for such studies due to the high sensitivity even on very short timescales. Our observations are described in Section \ref{sec:obs}. We then present an assessment of systematic effects for variability measurements using ALMA simulated observations in Section \ref{sec:alma_sim}. Our results on source detection is presented in Section \ref{sec:source_detection}, followed by our variability analysis in Section \ref{sec:variability} including the finding of a strong flare and the overall variability in the sample. We finally present a summary and our conclusions in Section \ref{sec:conclusions}. 

\section{Observations and Data Reduction}\label{sec:obs}

The Kleinmann-Low Nebula, a dense molecular cloud core close to the Becklin-Neugbauer object (herafter BN/KL; \citealt{becklin1967, bally2011}) was observed with ALMA during Cycle 5 (program 2017.1.01313.S, PI: J. Forbrich) at 3 mm (90 -- 105 GHz) on 2017 December 22, 27, and 29. A total of 8 epochs of $\sim$1.2~h each towards a single pointing centred at $\alpha_{\rm J2000}=05^{\rm h}35^{\rm m}14\fs5$ and $\delta_{\rm J2000}=-05\degr22\arcmin30\farcs6$ were obtained using the array configuration C43-5 with an average of 48 antennas per epoch (12-m array) on baselines of 15-2517~m (see Table \ref{tab:obs_logs}), where the longest baselines are particularly relevant to mitigate the extended emission in the Orion Nebula. This phase center position is 1~arcmin NW of $\theta^1$~Ori~C, the O7 type star providing most of the photons ionizing the Orion Nebula. In order to  prioritize time on source, the observations were carried out in dual-polarization mode recording the XX and YY correlations that allow us to recover the Stokes I intensity maps (Stokes Q is also accessible but insufficient to obtain overall linear polarization without additional calibration). Four continuum spectral windows with bandwidths of 1.875~GHz were used and centred at 90.5, 92.5, 102.5, and 104.5~GHz, each one consisting of 32 channels of 62.5~MHz-width. These spectral windows were chosen to avoid the strong lines of CO and its isotopologues.

\begin{table*}
\centering
\caption{ALMA Cycle 5 observation logs.}
\label{tab:obs_logs}
\begin{threeparttable}
\begin{tabular}{c c c c l c}
\hline
Epoch \# & Starting Time & Number of Antennas & Time on Source & \multicolumn{1}{c}{Synthesized beam size $^a$} & Sensitivity (1$\sigma$ rms) \\
 & (2017/UTC) &     & (h) & \multicolumn{1}{c}{(arcsec$^2$ ; $\degr$)} & ($\mu$Jy~beam$^{-1}$) \\
(1) &  (2)     & (3) & (4) & \multicolumn{1}{c}{(5)}                    & (6)                   \\
\hline   
1 & Dec 22 / 00:37:31  & 49 & 1.16  &  $0.41\times0.24$ ; 75 & 145\\
2 & Dec 22 / 01:58:28  & 49 & 1.16  &  $0.35\times0.25$ ; 84 & 118\\
3 & Dec 22 / 03:19:41  & 49 & 1.16  &  $0.32\times0.25$ ; 88 & 167\\
4 & Dec 22 / 04:40:49  & 49 & 1.16  &  $0.34\times0.25$ ; $-$86 & 274\\
5 & Dec 27 / 04:55:28  & 46 & 1.16  &  $0.36\times0.24$ ; $-$81 & 184\\
6 & Dec 29 / 02:52:24  & 46 & 1.16  &  $0.32\times0.24$ ; $-$86 & 249\\
7 & Dec 29 / 04:13:51  & 46 & 1.16  &  $0.34\times0.24$ ; $-$82 & 307\\
8 & Dec 29 / 05:35:50  & 46 & 1.17  &  $0.44\times0.23$ ; $-$78 & 291\\
 \hline
---& Concatenated image &--- & 9.3  &  $0.35\times0.24$ ; $-$88 & 42\\
\hline
\end{tabular}
\centering
\begin{tablenotes}
\item[Note] The array configuration used for all the observations was C43-5.

\item $^a$ Synthesized beam properties: $(\theta_{\mathrm{max}}\times\theta_{\mathrm{min}}$ ; PA)
\end{tablenotes}
\end{threeparttable}
\end{table*}

\begin{figure*}
    \centering
	\includegraphics[width=\linewidth]{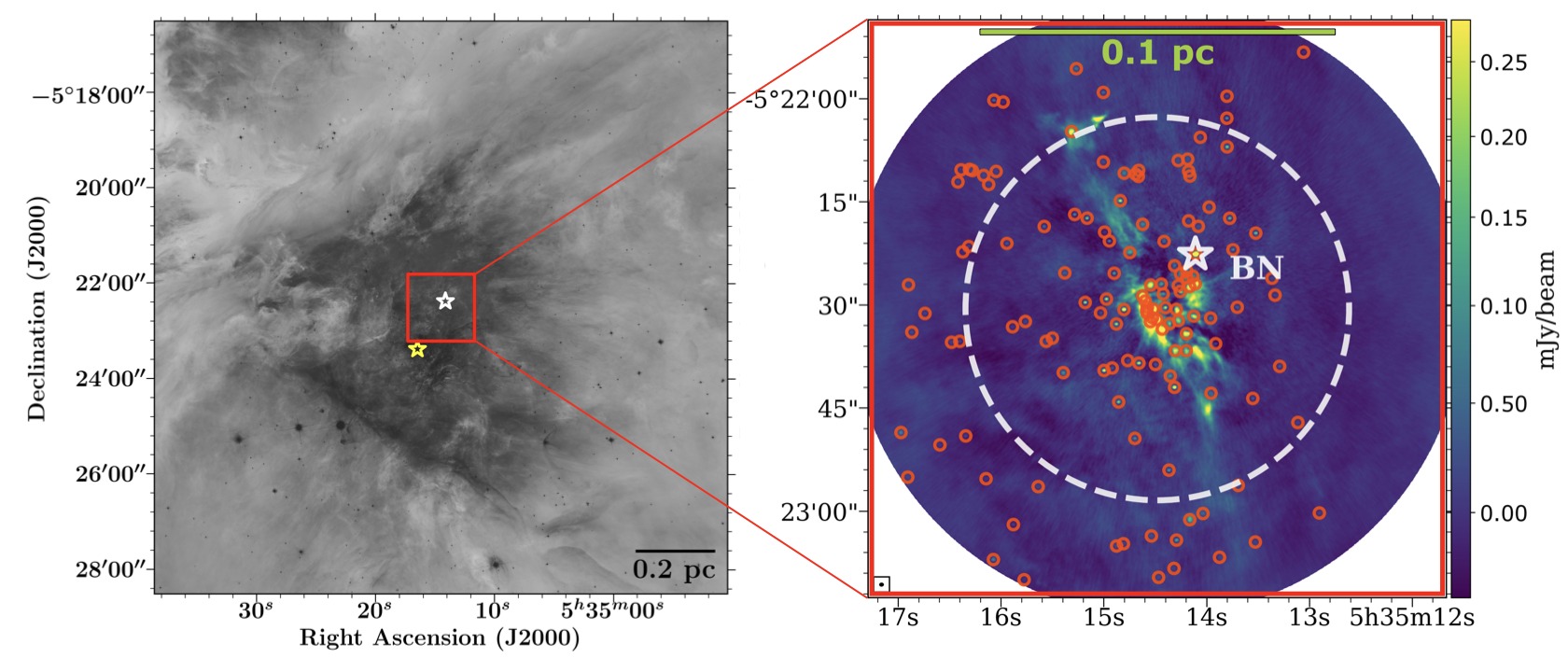}
    \caption{(left) HST r-band image (ACS/WFC) of the Orion Nebula with a field-of-view of 12$\times$12~arcmin$^2$ centred at the Orion-KL region (white star symbol) indicating in red the ALMA observed field. The yellow symbol indicates the position of $\theta^1$ Ori C in the Trapezium cluster (Background image credits: NASA, ESA, M. Robberto, and the Hubble Space Telescope Orion Treasury Project Team). (right) ALMA 3 mm continuum map of the Orion-KL region using the concatenated data (8 epochs combined). The white dashed circle indicates the HPBW at the central frequency ($\sim$0.93~arcmin) and the red circles indicate the 133 detected sources ($\geq5\sigma$). The white star symbol indicates the position of source BN as reference.}
    \label{fig:obs}
\end{figure*}

We used the pipeline-calibrated ALMA visibilities processed using the {\sc CASA}\footnote{Common Astronomy Software Application \citep{mcm07}.} software (release 5.4.1). The initial amplitude and bandpass calibrator was the quasar J0423-0120 and then the phase calibrator was J0529-0519, observed every 3 science scans (every 10~min). The calibrated dataset was imaged with the {\sc TCLEAN} task in {\sc CASA}. We used the Stokes plane $I$ and spectral definition mode `mfs' (Multi-Frequency Synthesis). The Hogbom deconvolution algorithm and a Briggs weighting method with a robustness parameter of 0.5  were used. The image size for all the observations is $2048\times2048$~pixels with a pixel size of 0.05~arcseconds and the mean synthesized beam size between all the individual maps is $0.36\times0.24$~arcsec$^2$, equivalent to physical lengths of 96$-$144~au at the distance of the ONC (assumed to be $\sim400$~pc; \citealt{grossschedl2018, kuhn2019}). Photometry was extracted from images that were corrected for the primary beam (PB) response following a PB gain level cut-off of 20~per~cent (pblim$=0.2$) and thus masking the image outside of a radius of $\sim$0.75~arcmin from the phase centre where the PB gain level reaches 20~per~cent due to PB attenuation. The resulting images cover a circular field of view of $\sim$1.5~arcmin in diameter with a half power beam width (HPBW) at the central frequency of $\sim$0.93~arcmin. (see Figure \ref{fig:obs}). 
In order to further reduce the impact of extended emission on the point-like source extraction process, we applied spatial filtering of the visibility data using baselines of the ($u$, $\nu$) range longer than 138 k$\lambda$ ($\sim$414~km) and therefore filtering out structures larger than $\sim$1.5~arcsec, on a field where the largest sources have sizes of $\lesssim$1~arcsec (excluding the extended component of the OMC1 hot core)\footnote{Interferometric observations are sensitive to a range of angular scales ($\theta_{res}$) given the range of antenna baselines as $\lambda/B_{max}<\theta_{res}<\lambda/B_{min}$, where $B_{min}$ and $B_{max}$ are the shortest and longest baselines, respectively.}.
Furthermore, with the angular resolution achieved by our observations and their corresponding physical lengths at the distance of the ONC, we are expecting unresolved emission from protostellar flares, magnetospheres, and even from larger magnetised structures confined within smaller scales than the synthesized beam corresponding to $\sim$100~au \citep{massi2008, salter2010}, while at the same time we are not expecting variability from extended structure (associated with thermal dust emission) on the short timescales studied here of minutes to days.

The overall sensitivity ranges between 0.18 and 0.31~mJy~beam$^{-1}$ among the different epochs where the highest rms noise levels are found in observations on the same day (29 December). These increased rms noise levels were due to poor weather conditions over the course of the last day of observations and also for epoch 4. The resulting image parameters for all the 1.16-h observations (hereafter ``epochs'') are summarized in Table \ref{tab:obs_logs}. An additional map with the concatenated data was generated following the same imaging procedure and spatial filtering used for the individual epochs. The improved sensitivity of the concatenated image reaches an rms noise of 0.04~mJy~beam$^{-1}$, corresponding to 36~per~cent of the noise levels of the individual epochs where weather conditions were better and 14~per~cent of the noise level in the epoch with the worst conditions (epoch 7). 
This combined image was used as a reference for source detection and to obtain the averaged peak flux densities reported in Table \ref{tab:catalog} and described in the following section. The premise here is that this deep image would yield the best source catalog as long as many sources have quiescent emission -- which is not always the case, as we will see below.

An additional set of images was generated at 20 and 4~minutes time resolution following the same procedure described above and aimed to look at the shorter timescales of the specific flare-like features found in the 1~h light curves described in section \ref{sec:variability}. Given the complex emission in this region, such time-sliced imaging and subsequent source fitting was used to obtain source photometry. The 20 minutes time resolution maps were chosen to include exactly 6 continuous science scans from the observations leaving outside only 2 remaining scans at the end of each individual epoch and equivalent to 3~minutes of observation. These 2 scans were then recovered when imaging the 4~minutes time resolution maps. The 20 minutes time resolution maps include $\sim$1.3~minutes of time dedicated to calibrations. All these images were used to generate light curves (LCs) at 1-h and 20-min time resolution, leading to 8 and 24 individual images, respectively. The rms noise levels for the 20-min images have values of 290 to 360$\mu$Jy~beam$^{-1}$. Only 17 images at 4~min time resolution were generated for the time intervals around the flare-like features of the most variable sources. Finally, given the discovery of a strong flare discussed in Section \ref{sec:orbs}, a set of 8-seconds time resolution images were generated only for a time interval of 40 minutes around the strong event following the same imaging procedure already described, resulting in 265 high-time resolution images with typical rms noise leves of $\sim$1.0~mJy~beam$^{-1}$.

\section{Simulated observations to assess systematic artificial variability in a complex region}\label{sec:alma_sim}

Our observations show evidence of widespread YSO millimeter variability on a wide range of timescales from minutes to days, including a strong flare. Our main focus is to characterise the strongest events that we can find in the resulting sample of sources, however this widespread variability extends down to the lowest measurable levels. An assessment of lower variability levels in this observations involves dealing with technical difficulties due to the ubiquitous complex multi angular scale emission in the ONC on top of its source density which necessarily requires time-sliced imaging of the whole field containing both resolved and unresolved sources and a constantly changing shape and size of the synthesized beam throughout the observations resulting in a variable background. This time-sliced imaging method needs to be applied to re-image the field at any time resolution followed by standard photometry to obtain flux measurements. Due to these complications it is expected that systematic effects will have an impact on flux measurements of unresolved sources and ultimately affecting variability measurements. The opposite scenario would be an isolated and unresolved source on top of a flat background where instead of time-sliced imaging a direct fit of a point source model to the visibilities would be a suitable method for flux measurements. 

In order to quantify the systematic effects described above and to determine what is the minimum level of variability in ONC sources that can be studied using our method we performed an analysis of simulated ALMA observations for a set of artificial, constant sources. Both the simulated observations and the artificial source properties reproduced as closely as possible our actual ALMA observations. These simulations consisted of 300 input images with a single artificial source in each, but using different source properties (brightness and shape) and different background properties as well. We made use of the {\sc Simobserve} task in CASA to first simulate the visibilities consisting of 7 observations of 1~h integration time each all of them at different hour angles ranging from $-5$~h to $+1$~h pointing towards the same phase center used in our actual observations as well as the same antenna configuration and reference date of the observations. We then made use of the {\sc Simanalyze} task to image the simulated visibilities. Finally, flux measurements and variability analysis were performed following the same method used for the analysis of the actual observations.

The artificial sources were 2D Gaussian models with a range of sizes for both major and minor axis equivalent to FWHM between 0.1 and 0.9~arcsec to include completely unresolved, marginally resolved and resolved sources in the experiment. The amplitude of these model sources were set to cover a range of peak flux density between 5 and 100~mJy~beam$^{-1}$. This set of basic parameters resulted in 100 initial artificial sources that were combined with three different background images taken directly from the actual observations using the concatenated data and within the HPBW primary beam. These three background sections of the concatenated image were chosen to represent three arbitrary levels of complexity from standard (largely clear) to highly complex (contaminating extended emission) with rms noise levels ranging between 0.1 and 2.5~mJy~beam$^{-1}$. Each source was located at the center of the three different background images. This resulted in 300 input sky models whose visibilities were simulated for 7 different hour angles (HAs) on the sky and subsequently imaged with {\sc Simanalyze} with a pixel size of 0.05~arcsec and an image size of 100~pixels per side. This results in a total of 2100 simulated images (seven observations per each of the 300 input sky models). 

\begin{figure*}
    \centering
	\includegraphics[width=.7\linewidth]{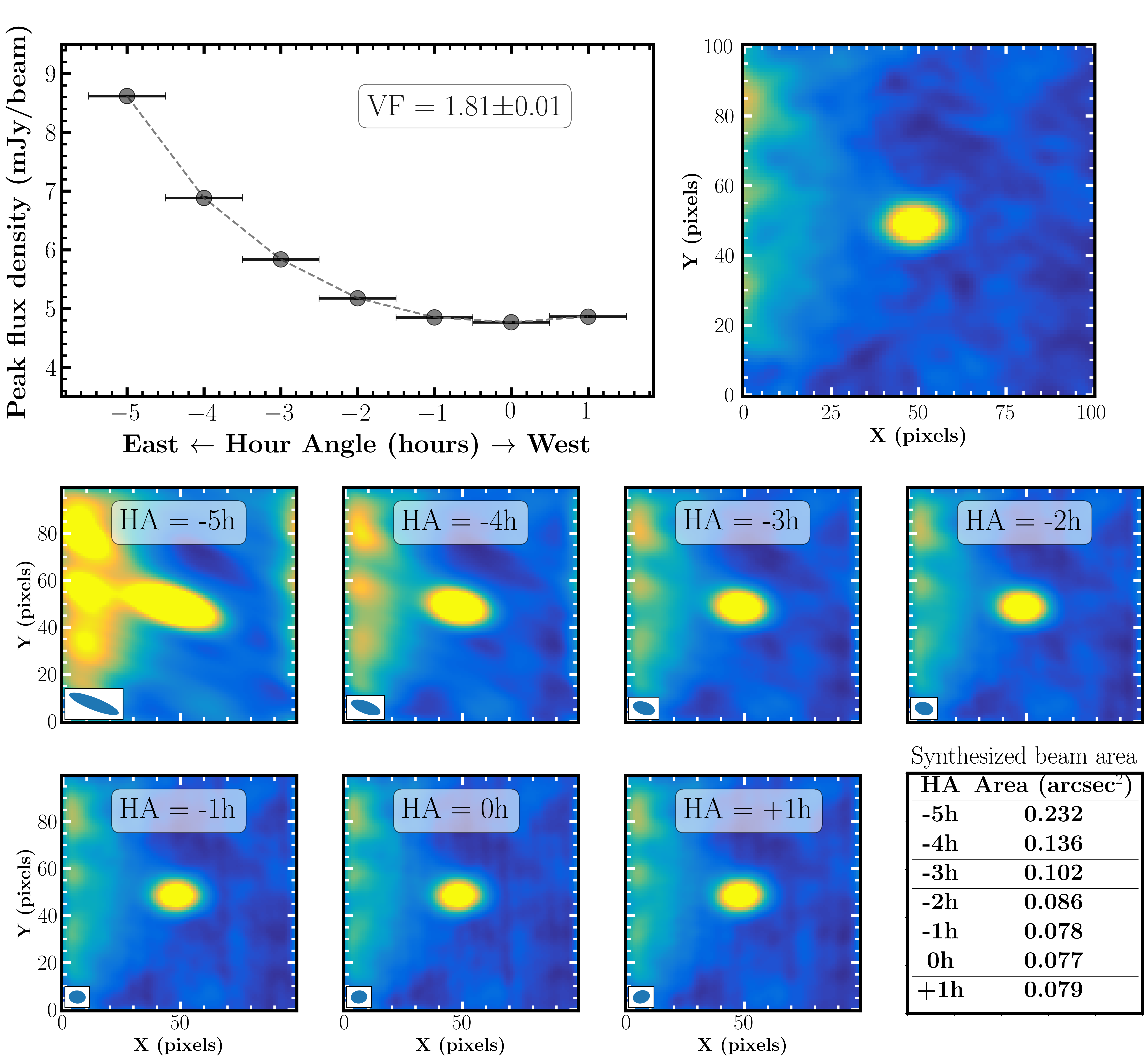}
    \caption{Simulated constant source and its resulting artificial variability. Left-hand panel in the top row show the resulting LC and VF. Right-hand panel in the top shows the input image containing the artificial source already combined with the background image. The middle and bottom rows show the resulting simulated observations labeled with the corresponding HA and with the resulting synthesized beam indicated in the lower-left corner of each map. The table in the bottom-right panel indicates the synthesized beam area for each observation.}
    \label{fig:sim_LC}
\end{figure*}

In order to extract the flux information from each of the 2100 simulated images, we applied the same method used for source detection in the actual data. Following the source extraction method described in \citet{vargas2021}, we obtained flux information using a Gaussian fitting algorithm based on the {\sc IMFIT} task in {\sc CASA} that iterates over each input source using different values for fitting area around the source and different offsets from the input position to avoid nearby contamination. The flux measurements were then used to analyze the LCs of each source in order to assess the maximum change in peak flux density throughout the 7 observations, hereafter variability factor (VF), and defined as the ratio between the maximum and minimum peak flux density in the LCs. Figure \ref{fig:sim_LC} shows an example for the resulting analysis and includes the LC of the source in the top left panel and its corresponding maximum change in peak flux density (VF=1.81$\pm$0.01). The sky model (shown in the top right panel) corresponds to the input image for the simulation and contains the artificial source combined with one of the three real background images. In this example, this background section of the concatenated data that has an rms noise of 0.5~mJy~beam$^{-1}$, the source model already combined with the background image has a peak flux density of $5.02\pm0.02$~mJy~beam$^{-1}$ and an original area of $\sim$0.26~arcsec$^2$. The resulting simulated observations shown in the middle and bottom panels include the resulting beam in the lower left corner of each map, already highlighting its changes in shape, size and orientation. The areas of these resulting beams are indicated in the bottom-right panel for each observed HA to illustrate how different elevations largely affect the synthesized beam subsequently affecting our final flux density measurements. 

As in the example shown in Figure \ref{fig:sim_LC}, the large range of source properties and the different backgrounds used are differently affected by this change in the synthesized beam over the course of the observations, and thus also resulting in a range of artificial variability levels. These artificial variability spans a range of 1.1$\leq$VF$\leq$2.1 and a median value of VF$=$1.51. 

These results represent a conservative and likely overestimated assessment of the maximum systematic VF we could find in our actual observations. This is because we have used a wider range of source properties and also a wider range of elevations than the actual ALMA observations of the ONC comprise, which had a maximum HA coverage between $-$3h20m and $+$3h13m, with resulting source elevations between $40\degr$ and $72\degr$, these simulations included a HA range of $-5$h$\leq$HA$\leq$+1h, which results in source elevations between $15\degr$ and $72\degr$. If we limit these simulations to $-$3h20m$<$HA$<+1$h, we then obtain a maximum systematic variability of VF$\sim$1.6. As a compromise between these considerations and the results from the full sample in the simulations, a reasonable cut-off for systematic variability that can affect at least a sample of sources given certain conditions (e.g., source size in combination with a complex background) is VF$=$2. In these simulations, $\sim$76~per~cent of the sources show VF$\leq$2, $\sim$17~per~cent show 2.0$\leq$VF$\leq$2.1, and there is a 7~per~cent for which it was not possible to obtain a final VF since these were not detected when combined with the most complex background. While sources in our actual ALMA data with VF$\leq$2 may still show real variability, the main focus in our study is to find the strongest events and how often those occur rather than a detailed study of minor variability. Here, we discuss these highly variable sources as well as important considerations for the study of lower levels of variability for similar datasets. Also, while we have used a wide range of parameters for sources and backgrounds to determine a conservative lower limit in our variability analysis, these do not necessarily comprise the full range of scenarios for every single source in the actual observations and therefore such variability for specific sources would still need visual inspection. 

Since the systematic effects relevant here are linked to changes in synthesized beam and in turn this is linked to source elevation, then these systematic effects will generally not occur on arbitrary timescales. An example of this can be seen in the lightcurve shown in Figure \ref{fig:sim_LC} for which the artificial change in peak flux density smoothly develops with elevation. The only exception would be an adjacent contaminant that would pass through the beam as it rotates throughout the observations. Even in such a scenario the resulting effect would not compromise the detection of a short flare which will have different characteristics.

Beyond these systematic effects, which dominate our analysis, \citet{francis2020} analyzed the accuracy of ALMA flux calibration and the impact on variability searches, which is particularly relevant for isolated sources with a clean background. They find that with improved calibration strategies the uncertainty can be lowered to a few percent, but this is beyond what can be achieved in our complex target region.

\section{Results from ALMA observations}\label{sec:results}

\subsection{Source Detection}\label{sec:source_detection}

Compact source detection methods applied to radio maps towards crowded and complex star-forming regions face a challenge due to contamination from spatially filtered complex extended emission. Even after applying additional spatial filtering, this often remains as uneven noise with occasional spurious point-like emission, and therefore automated source extraction methods typically require significant manual intervention to deal with artifacts. We thus searched for compact sources by visual inspection of the concatenated image followed by an automatic search only on the position of known X-ray sources in the Chandra Orion Ultra-deep Project (COUP; \citealp{get05b}), known sources detected at cm-wavelengths with the VLA reported by \citet{for16} and \citet{vargas2021}, and millimeter sources reported in \citet{friedel2011}, \citet{eisner2016}, and \citet{otter2021}. These catalogues provide an updated and well characterized sample of X-ray and radio sources likely tracing the young stellar population in the Orion BN/KL region. Along with the multiwavelength tracers of young stars, variability itself, particularly at the heart of the OMC1 cloud, would most likely originate from a YSO and it is thus a suitable tool for source identification even for deeply embedded sources inaccessible at other wavelengths, particularly IR and optical, where a strong radio flare would be the only observable tracer in such a case. 

From our search in the aforementioned X-ray and radio surveys, within the HPBW primary beam (r$<$0.47~arcmin from the phase centre) there are 52 COUP sources.
Among these X-ray sources, 48~per~cent have a counterpart in our catalog using a search radius of 0.5~arcsec where only two additional nearby X-ray sources could be included with separations $\leq$0.7~arcsec if applying a search radius of 1~arcsec, one of them COUP 599, an unclear counterpart of source BN \citep{grosso2005}. 
Similarly, we detect 40~per~cent of the 58 VLA sources within the HPBW primary beam (6~cm observations with angular resolutions of $\sim$0.4~arcsec). 
Only 4 additional nearby VLA sources (angular separations $\leq$0.8~arcsec) can be included if extending the search radius from 0.5~arcsec to 1~arcsec \citep{for16, vargas2021}. 
On the other hand, we detect 89~per~cent of the 28 millimeter sources within the HPBW primary beam reported in \citet{friedel2011} based on 3~mm CARMA observations at different spatial resolutions down to a synthesized beam size of $\sim$0.5~arcsec.
Two of the remaining 3 ``non-detected'' millimeter sources appear as extended structures ($>$1.5~arcsec) in our ALMA observations (sources C2 and C30, see Table 1 in \citealt{friedel2011}) and are not included in our analysis. 
We also detect the 4 millimeter sources in the field reported in \citet{eisner2016} using 1.3~mm ALMA observations with angular resolution of $\sim$1~arcsec, listed as proplyds and detected in optical and/or near-IR bands \citep{ricci2008, hillenbrand2000}. 
Finally, in the ALMA millimeter survey presented in \citet{otter2021} there are 61 within the HPBW primary beam of which 77~per~cent have counterparts in our catalog within 0.1~arcsec following a search radius of 0.5~arcsec.
There are 14 millimeter sources from \citet{otter2021} not detected in our work that lie within the HPBW primary beam of our observations. Their reported 3-mm flux measurements are $\leq$0.4~mJy except for two sources with flux measurements of 0.673$\pm$0.010~mJy and 0.850$\pm$0.007~mJy, identified as sources 40 and 38 in their catalog, the former located $\sim$0.6~arcsec north-west from source BN, where the local rms noise level is $\sim$0.3~mJy~beam$^{-1}$ while the latter, located in an empty field with local rms noise of $\sim$0.04~mJy~beam$^{-1}$, would be expected to be clearly detected with S/N$>$5 if it was a constant source.

We obtained flux information following the source extraction method described in section \ref{sec:alma_sim} based on a Gaussian fitting algorithm using the {\sc IMFIT} task in {\sc CASA}. Due to the presence of noise peaks with S/N levels in the range of $\sim3-4$, we have enforced a detection threshold of 5$\sigma$, leading to a total of 133 sources. We noted a significant improvement for source detection by using the additional spatial filtering of the visibility data allowing a 39~per~cent increase in detected sources, amongst these the flaring source discussed in section \ref{sec:orbs} which is surrounded by complex emission that does not allow to fit a Gaussian component unless applying the additional spatial filtering. The main resulting parameters (position, peak flux densities, and source structure) for the 133 detected sources are listed in Table \ref{tab:catalog} and were obtained from the concatenated data (full catalogue available in the online version). 

All the detected sources are indicated by red symbols in the right panel of Figure \ref{fig:obs} overlaid on the ALMA 3~mm continuum map from the concatenated data. The continuum map shown in the background in Figure \ref{fig:obs} was generated without the additional spatial filtering described in section \ref{sec:obs} for illustrative reasons in order to highlight the complex extended emission particularly in the inner region. The source distribution shows a higher number density towards the eastern side of the cluster with no detections above $5\sigma$ in the westernmost area. A similar spatial distribution is found at cm-wavelengths as well as in the X-ray and NIR bands \citep{for16}. The lower source density at X-ray and NIR wavelengths can be associated with higher extinction levels, consistent with the higher dust emission towards the western region in the ONC as seen at submillimeter-wavelengths \citep{dif08}, while the radio population distribution, essentially unaffected by extinction, is likely tracing the actual YSO distribution with the exception of the intrinsically faint millimeter sources. 

The goal of this work is to search for short-term millimeter variability from minutes to days associated with nonthermal radio emission in protostars, and therefore we do not intend to study disk properties here, which have been discussed in detail elsewhere (e.g., \citealt{eisner2016, eisner2018, otter2021}). In this context, if the measured flux is dominated by disk emission we will not expect any short-timescale variability, however the resulting flux measurements from insufficiently resolved or totally unresolved sources are likely to be a combination of both the disk component and flares and thus it becomes a relevant concern for disk mass studies. In this regard, while COUP sources in our sample already represent our best tracer of the young stellar population including objects associated with disks, we still searched for counterparts in multiwavelength surveys in Orion as an additional approach to identify the fraction of known protoplanetary disks in our sample and quantify to what extent millimeter flares could potentially dominate such emission towards these sources.

The typical size for circumstellar disks from optical studies within our observed area of the ONC is $\sim$130~au with just a few larger than 150~au \citep{vicente2005}. The spatial resolution in our observations is equivalent to spatial scales of $\sim$140~au at the distance of the ONC, and we are therefore looking at unresolved or just marginally resolved protoplanetary disks in the region. We searched for protoplanetary disk counterparts in the literature within 1~arcsec to account for the combined uncertainties between different observations and for the emission scales at optical and/or infrared wavelengths of these systems that could still be associated with a millimeter counterpart within this search radius.
Within this field there are 21 out of the 162 protoplanetary disks reported at optical wavelengths in \citet{vicente2005} whereas we find 14 mm-counterparts in our catalog with separations between 0.3 $-$ 0.8~arcsec. Based on similar observations with the HST, \citet{ricci2008} reported 29 protoplanetary disks within this field, while we find 16 mm-counterparts in our catalog with maximum separations of $\sim$0.5~arcsec. In addition to the four 1.3-mm sources from \citet{eisner2016} mentioned at the beginning of this section, 24 out of 29 sources detected at 0.85~mm that fall within our observed field are detected in our observations \citep{eisner2018}. At least $\sim$25~per~cent of well characterized disks are associated with sources in our sample, of which $\sim$73~per~cent of them are already COUP counterparts, and we will also be able to assess any variability associated with these systems. If we include the sample of small protoplanetary disks studied in \citet{otter2021} there is then a fraction $\sim$66~per~cent of characterized disks in our sample of which $\sim$64~per~cent are COUP sources.

After identifying the fraction of sources associated with well-characterized protostellar systems and a large sample of mm-sources with multiwavelength properties characteristic of young stellar objects we then performed a systematic search for variability based on the 1~h epochs spanning more than a week of observations which we describe in the following section.

\begin{landscape}
\begin{table}
\centering
\small
\caption{ALMA 3 mm catalogue: Source properties and variability measurements in the Orion-KL region.}
\label{tab:catalog}
\begin{threeparttable}
\tabcolsep=0.1cm
\begin{tabular}{cccrcccccccccc}
\hline
                    & &          &     &    Deconvolved Size   &\multicolumn{2}{c}{1~h time resolution}&\multicolumn{2}{c}{20~min time resolution}& & & & \\
                    \cmidrule(lr){6-7} \cmidrule(lr){8-9} 
$\alpha(2000)$      & $\delta(2000)$& ID &\multicolumn{1}{c}{Peak Flux Density}& $\theta_{\mathrm{max}}\times\theta_{\mathrm{min}}$ ; P.A.  & COUP & F16$^a$ & V21$^a$ & O21$^a$  & Additional\\
$({\rm ^h\,^m\,^s})$& $(\degr\,\arcmin\,\arcsec)$&    &\multicolumn{1}{c}{(mJy bm$^{-1}$)}& (arcsec$^2$ ; $\degr$) & & &  & & counterpats$^b$\\
(1) & (2)  &  (3)  & \multicolumn{1}{c}{(4)}  & (5) & (6) & (7) & (8) & (9) & (10)\\
\hline
  05:35:12.9031 $\pm$ 0.0067  & -5:23:00.2701 $\pm$ 0.0028  &  1   & 0.554  $\pm$ 0.020  &                                    &        &         &        &  80   &   \\      
  05:35:13.0633 $\pm$ 0.0181  & -5:21:53.2754 $\pm$ 0.0051  &  2   & 0.734  $\pm$ 0.037  & 0.53 $\times$ 0.08 ;  102 $\pm$ 2  &  516   &         &        &  77   &   \\   
  05:35:13.1107 $\pm$ 0.0068  & -5:22:47.1013 $\pm$ 0.0075  &  3   & 0.274  $\pm$ 0.015  &                                    &  524   &  107    &  128   &       & 131-247  \\   
  05:35:13.2887 $\pm$ 0.0054  & -5:22:38.9703 $\pm$ 0.0028  &  4   & 0.703  $\pm$ 0.023  &                                    &  539   &         &        &  13   &   \\   
  05:35:13.3331 $\pm$ 0.0361  & -5:22:28.5864 $\pm$ 0.0113  &  5   & 0.195  $\pm$ 0.020  &                                    &        &         &        &       &   \\   
  05:35:13.3655 $\pm$ 0.0112  & -5:22:26.1496 $\pm$ 0.0035  &  6   & 0.296  $\pm$ 0.017  &                                    &  538   &  117    &        &  36   &   \\   
  05:35:13.5207 $\pm$ 0.002   & -5:22:19.5594 $\pm$ 0.0011  &  7   & 2.491  $\pm$ 0.027  & 0.13 $\times$ 0.11 ;  79 $\pm$ 80  &  551   &  127$^*$&  135   &  46   & 135-220  \\   
  05:35:13.5242 $\pm$ 0.0085  & -5:23:04.4708 $\pm$ 0.0061  &  8   & 0.405  $\pm$ 0.024  &                                    &  552   &         &        &  107  & HC360  \\   
  05:35:13.5497 $\pm$ 0.0042  & -5:22:43.618  $\pm$ 0.0027  &  9   & 0.624  $\pm$ 0.016  &                                    &        &         &        &  7    &   \\   
  05:35:13.6895 $\pm$ 0.0066  & -5:22:56.2309 $\pm$ 0.007   &  10  & 0.175  $\pm$ 0.013  &                                    &  563   &         &        &  1    &   \\   
  05:35:13.7027 $\pm$ 0.0068  & -5:22:30.3492 $\pm$ 0.0034  &  11  & 0.482  $\pm$ 0.017  &                                    &  574   &         &        &  29   &   \\   
  05:35:13.7428 $\pm$ 0.0031  & -5:22:21.9904 $\pm$ 0.0017  &  12  & 0.955  $\pm$ 0.017  &                                    &  573   &         &        &  41   & 137-222  \\   
  05:35:13.7756 $\pm$ 0.0016  & -5:22:17.399  $\pm$ 0.0008  &  13  & 2.095  $\pm$ 0.018  & 0.13 $\times$ 0.03 ;  123 $\pm$ 7  &  572   &  139    &        &  48   &   \\   
  05:35:13.8003 $\pm$ 0.0011  & -5:22:07.0349 $\pm$ 0.0007  &  14  & 3.199  $\pm$ 0.022  & 0.12 $\times$ 0.04 ;  174 $\pm$ 7  &  579   &  142    &  142   &  59   & 138-207  \\   
  05:35:13.8025 $\pm$ 0.0051  & -5:21:59.6729 $\pm$ 0.0028  &  15  & 0.802  $\pm$ 0.026  &                                    &        &  140    &        &  78   &   \\   
  05:35:13.8032 $\pm$ 0.0019  & -5:22:02.8587 $\pm$ 0.0011  &  16  & 2.583  $\pm$ 0.030  &                                    &  578   &         &        &  62   &   \\   
  05:35:13.8694 $\pm$ 0.0081  & -5:23:06.714  $\pm$ 0.0039  &  17  & 0.345  $\pm$ 0.014  & 0.23 $\times$ 0.05 ;  61 $\pm$ 12  &        &         &        &  106  &   \\   
  05:35:13.9099 $\pm$ 0.0066  & -5:22:35.632  $\pm$ 0.0032  &  18  & 0.546  $\pm$ 0.020  &                                    &  591   &         &        &  16   &   \\   
  05:35:13.9597 $\pm$ 0.0249  & -5:22:42.8081 $\pm$ 0.0186  &  19  & 0.263  $\pm$ 0.021  & 0.61 $\times$ 0.12 ;  50 $\pm$ 5   &        &         &        &       &   \\   
  05:35:13.9611 $\pm$ 0.0135  & -5:22:31.9325 $\pm$ 0.0073  &  20  & 0.479  $\pm$ 0.036  &                                    &  590   &         &        &  23   &   \\   
  05:35:13.9733 $\pm$ 0.0137  & -5:22:15.7734 $\pm$ 0.0094  &  21  & 0.196  $\pm$ 0.025  &                                    &        &         &        &       &   \\   
  05:35:14.0330 $\pm$ 0.0064  & -5:23:00.3572 $\pm$ 0.0038  &  22  & 0.594  $\pm$ 0.026  &                                    &        &         &        &  81   &   \\   
  05:35:14.0600 $\pm$ 0.0195  & -5:22:05.6447 $\pm$ 0.0094  &  23  & 0.268  $\pm$ 0.028  & 0.12 $\times$ 0.05 ;  70 $\pm$ 37  &        &         &        &  60   &   \\   
  05:35:14.0820 $\pm$ 0.0167  & -5:22:18.5331 $\pm$ 0.0074  &  24  & 0.259  $\pm$ 0.028  &                                    &        &         &        &       &   \\   
  05:35:14.1055 $\pm$ 0.0005  & -5:22:22.6444 $\pm$ 0.0002  &  25  & 70.991 $\pm$ 0.183  & 0.07 $\times$ 0.03 ;  9 $\pm$ 9    &  599   &  162    &  156   &  39   &  source BN  \\   
  05:35:14.1080 $\pm$ 0.0261  & -5:22:26.9292 $\pm$ 0.0134  &  26  & 1.719  $\pm$ 0.106  &                                    &        &         &        &       &   \\   
  05:35:14.1242 $\pm$ 0.0096  & -5:22:31.6324 $\pm$ 0.0042  &  27  & 1.814  $\pm$ 0.067  & 0.37 $\times$ 0.10 ;  68 $\pm$ 5   &        &         &        &       &   \\   
  05:35:14.1319 $\pm$ 0.0281  & -5:22:25.8623 $\pm$ 0.0268  &  28  & 0.535  $\pm$ 0.048  & 0.75 $\times$ 0.17 ;  43 $\pm$ 4   &        &         &        &       &   \\   
  05:35:14.1642 $\pm$ 0.0135  & -5:22:11.2622 $\pm$ 0.0142  &  29  & 0.142  $\pm$ 0.021  &                                    &        &         &        &       &   \\   
  05:35:14.1656 $\pm$ 0.0052  & -5:23:01.2915 $\pm$ 0.0225  &  30  & 1.318  $\pm$ 0.046  & 1.23 $\times$ 0.16 ;  174 $\pm$ 1  &        &  166    &  159   &       & 142-301  \\
\hline
\end{tabular}
\centering
\begin{tablenotes}
\item $^a$ Source identificatins for counterparts in: F16 \citet{for16}, V21 \citet{vargas2021}, and O21 \citet{otter2021}. Source from \citet{for16} reported as nonthermal centimeter counterpart in the VLBA follow-up \citep{forbrich2021} are marked with an asterisk symbol in column (11).
\item $^b$ Additional counterparts associated with known circumstellar disks: \citealt{ricci2008, eisner2016, eisner2018, vicente2005, hillenbrand2000}.
\item The full catalog with 133 listed objects is available as supplementary material.
\end{tablenotes}
\end{threeparttable}
\end{table}
\end{landscape}


\subsection{Radio Variability}\label{sec:variability}

As stated above, we here aim to search for the occurrence rate of the strongest short-term variability. Employing the same source extraction method used on the concatenated data, we search for emission toward all the sources in our catalog (Table \ref{tab:catalog}) in all the observed 1-h epochs. 56~per~cent of the sources were detected in all the individual epochs, 90 per cent were detected in at least half of them and every source was detected at least once. While non-detections can be explained by the increased rms noise levels in the individual images of up to 0.3~mJy~beam$^{-1}$ compared to the 0.04~mJy~beam$^{-1}$ in the concatenated image (see column 6 in Table \ref{tab:obs_logs}), this could still suggest some degree of variability for sources that would only be detected during a flare.

The peak flux measurements from the individual epochs were used to generate the LCs at 1~h time resolution for all the sources as shown for source 74 (the ORBS) in the left-hand panel in Figure \ref{fig:LC_ORBS}. This LC covers a strong flare discussed in detail in the following section. If no peak above a 5$\sigma$ detection threshold was found, then three times the local rms noise was used as an upper limit (red symbols in Figure \ref{fig:LC_ORBS}). The maximum change in peak flux density is defined as VF (see section \ref{sec:alma_sim}). When an upper limit is used as a minimum, then the VF is reported as a lower limit for variability since we are not able to account for the true peak flux density during that minimum. The same criteria were used to generate the LCs at different time resolution as shown in the middle and right-hand panel of Figure \ref{fig:LC_ORBS} with LCs at 20- and 4-min time resolution.


\subsubsection{Strong Flaring source ORBS}\label{sec:orbs}

\begin{figure*}
    \centering
	\includegraphics[width=\linewidth]{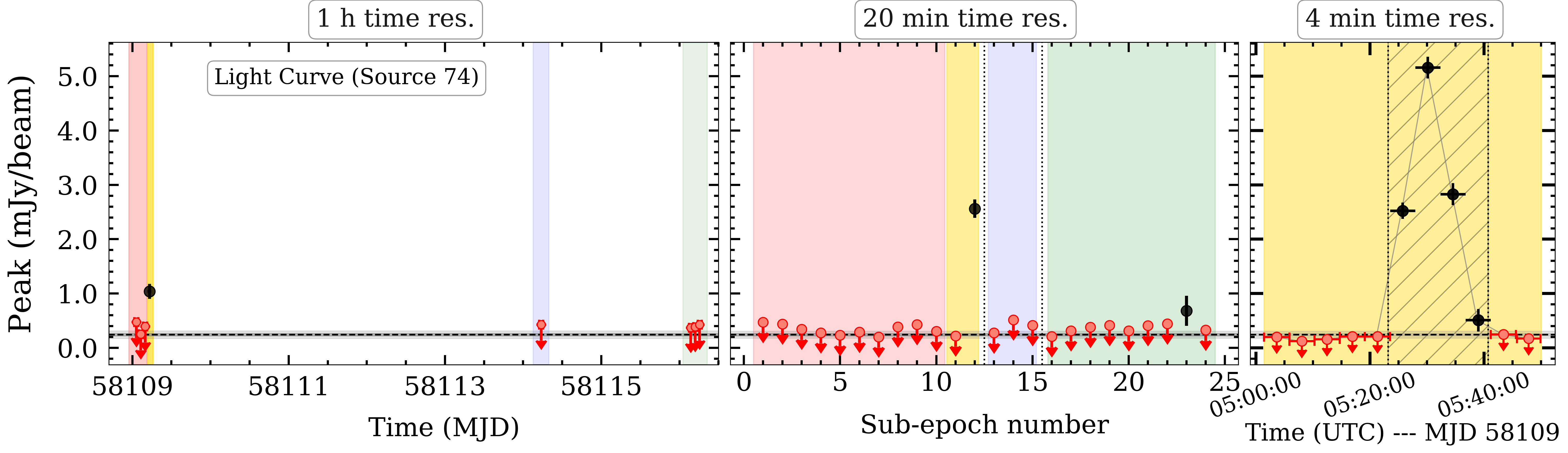}
    \caption{Radio LCs of source 74 (ORBS) at 1-h (left-hand panel), 20-min (middle panel), and 4-min (right-hand panel) time resolutions. The middle panel indicates the horizontal axis in arbitrary units representing consecutive epochs with their corresponding time intervals highlighted in red, blue, green, and yellow areas as shown in the left-hand panel. The yellow area indicates the time interval around the flare event then highlighted in the following two panels at higher time resolution. Detections are shown in black with 3$\sigma$ error bars. Upper limits are indicated in red (three times the local rms noise). The dashed horizontal line represents the averaged peak flux density from the concatenated data with 1$\sigma$ and 3$\sigma$ error bands in grey. The dashed background in the 4-min LC spans the time interval shown in Figure \ref{fig:LC_ORBS_8s}.}
    \label{fig:LC_ORBS}
\end{figure*}

\begin{figure}
    \centering
	\includegraphics[width=\linewidth]{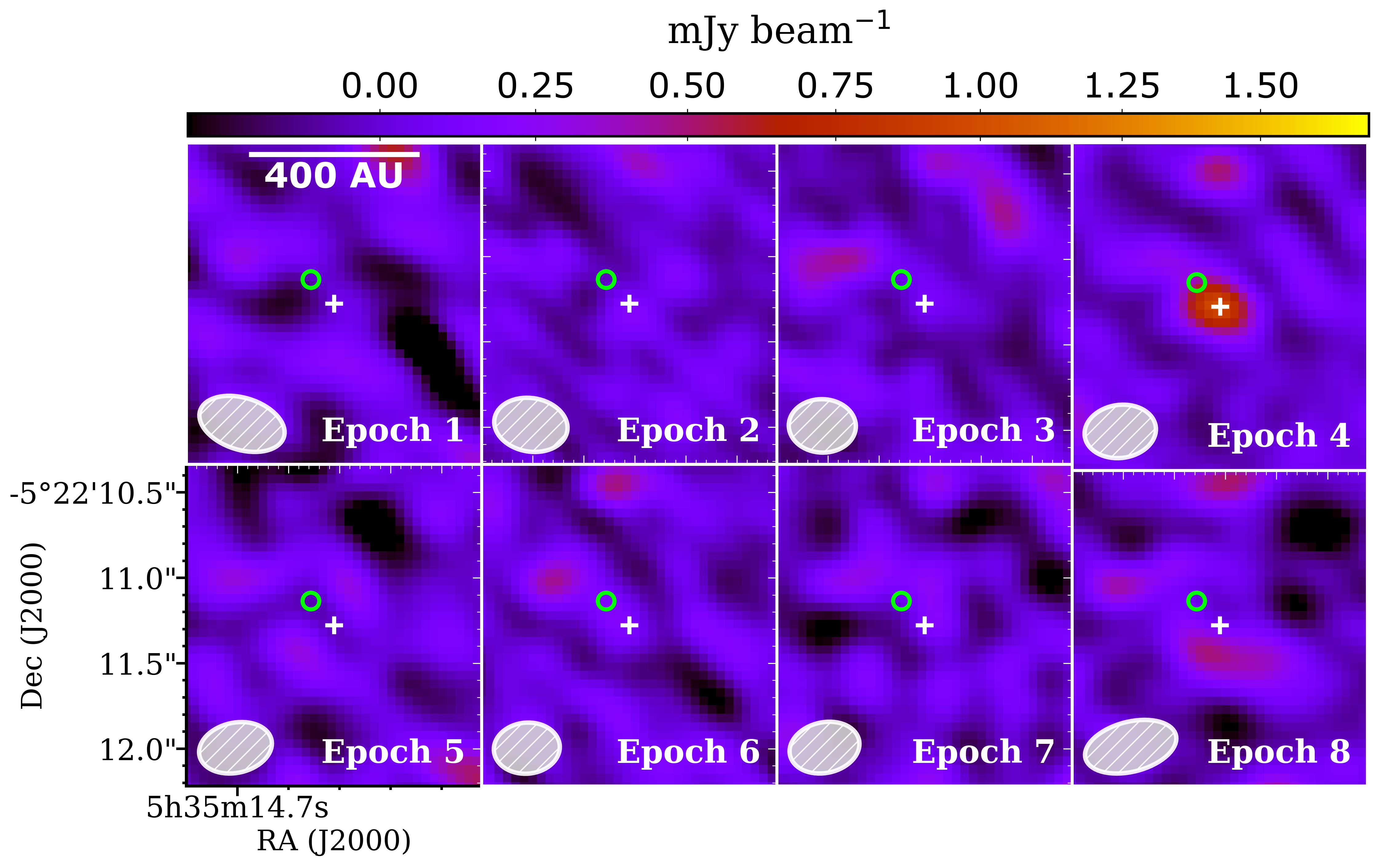}
    \caption{Continuum maps from the eight individual epochs at 1-h time resolution (listed in Table \ref{tab:obs_logs}) around the position of the flaring source ORBS. The green circles indicate the position of the X-ray source COUP 647 while the white plus symbol indicates the position of the millimeter detection in epoch 4.}
    \label{fig:orbs}
\end{figure}

\begin{figure*}
    \centering
	\includegraphics[width=\linewidth]{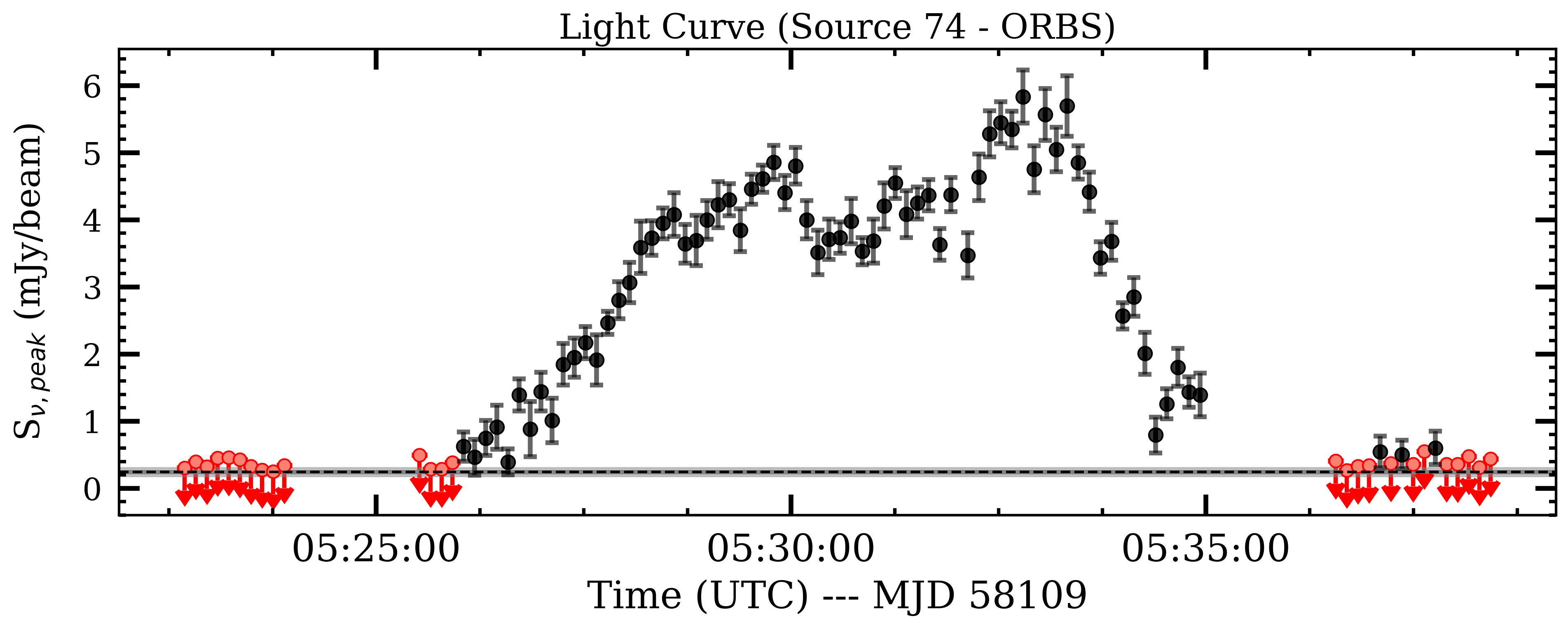}
    \caption{Radio LC of source ORBS at 8-s time resolution following same symbol notation from Figure \ref{fig:LC_ORBS}.}
    \label{fig:LC_ORBS_8s}
\end{figure*}

Visual inspection of the individual eight epochs (see Table \ref{tab:obs_logs}) for a subset of sources in our catalog led to the discovery of a flare object identified as source 74 in our catalogue (Table \ref{tab:catalog}). It only appears as a very faint source with an average peak flux density of $0.241\pm0.022$~mJy~beam$^{-1}$ in the concatenated image where it is only possible to fit a Gaussian component following the additional spatial filtering of the visibility data described in Section \ref{sec:obs}. Otherwise this source would not be detected, mostly due to complex emission in the surrounding area, the presence of a nearby source within $\sim$0.4~arcsec (source 75 in our catalogue) and its relatively faint average peak flux density. Based on the eight individual 1-h epochs, this source was only detected once, remaining undetected for more than three hours since the start of the observations to then peak at $1.039\pm0.046$~mJy~beam$^{-1}$ (S/N$\sim$23) in epoch 4. It remained undetected in the following epochs five days later with an average 3$\sigma$ upper limit of $\sim$0.4~mJy~beam$^{-1}$ (see left-hand panel in Figure \ref{fig:LC_ORBS} and the corresponding continuum maps in Figure \ref{fig:orbs} for the individual epochs). This led to a lower limit variability with VF$\sim$4.1 within $\sim$2.7~h against epoch 2. Its LC at 20~min time resolution (see middle panel in Figure \ref{fig:LC_ORBS}) allows us to further constrain this event to develop in less than an hour with a significantly increased VF$\sim$21$\pm$4 and a peak flux density of $2.562\pm0.056$~mJy~beam$^{-1}$. This source clearly was only detected during this flare which was bright enough to still allow its (faint) detection in the averaged 8-h image.

Its high S/N even at shorter timescales allows us to constrain the development of this event for which we generated its 4-min and then 8-s time resolution LCs shown in the right-hand panel in Figure \ref{fig:LC_ORBS} and separately in Figure \ref{fig:LC_ORBS_8s}, respectively. The 8-s images span a time interval of 40~min around the flare event, however, the LC in Figure \ref{fig:LC_ORBS_8s} is only displaying an interval of $\sim$17~min (96 images in total). Outside this time interval there are no detected peaks above a 5$\sigma$ threshold.

\begin{figure}
    \centering
	\includegraphics[width=\linewidth]{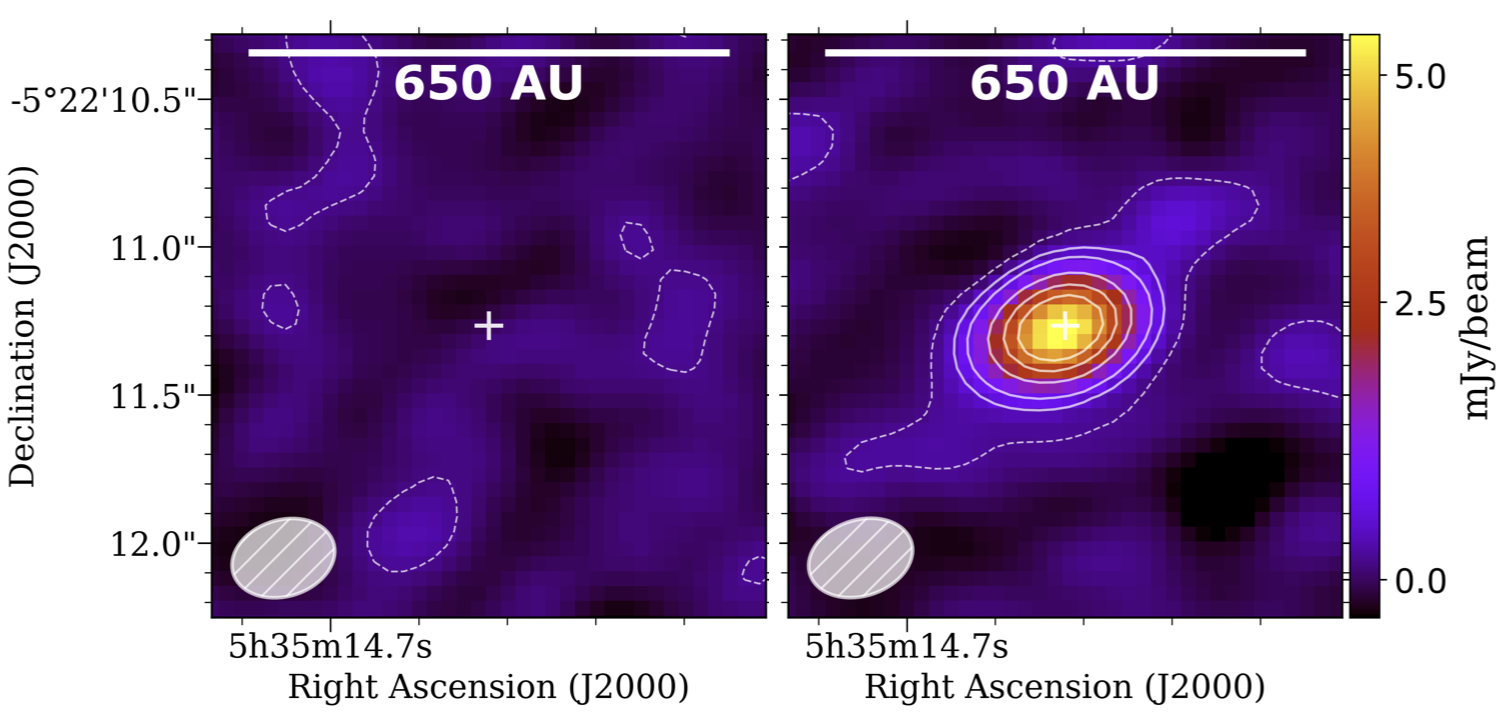}
    \caption{Continuum maps of the 8-s time resolution images at the position of the flaring source ORBS. The left-hand panel corresponds to a time frame just prior to its first detection at 05:26 UTC (third upper limit after 5:25 UTC in Figure \ref{fig:LC_ORBS_8s}), and the right-hand panel shows the maximum peak at 05:32:43 UTC. Contour levels are 1$\sigma$, 3$\sigma$, 5$\sigma$s, 10$\sigma$, 15$\sigma$, and 20$\sigma$ rms levels. The plus symbol indicates the position of the peak from the right panel.}
    \label{fig:ORBS_max_vs_min}
\end{figure}

While at these two time resolutions we are already constraining the brightness of the event with similar peak flux density of $5.159\pm0.066$~mJy~beam$^{-1}$ at 4-min time resolution and a maximum of $5.835\pm0.132$~mJy~beam$^{-1}$ at 8-s time resolution, it is only at the highest 8-s time resolution that a more detailed substructure in the LC is seen allowing us to constrain the flare duration to $\sim$10~minutes with a rise time of $\sim$4~minutes from the first detection (05:26 UTC) until the first peak corresponding to an order of magnitude change in peak flux density in such a short timescale. However, the presence of several features in the LC may not necessarily correspond to the same event and the flare duration refers to the entire event in the LC which shows a brief decline at 05:30 UTC generating two main peaks, the second one $\sim$7 minutes after the first detection and just $\sim$3~min from the first peak. The second peak is the maximum already mentioned and the first one just slightly fainter at $4.856\pm0.085$~mJy~beam$^{-1}$. Figure \ref{fig:ORBS_max_vs_min} shows the 8-s time resolution continuum maps in a time frame just prior to the first detection of the source (left-hand panel) and at the maximum peak (right-hand panel) seen in the LC shown in Figure \ref{fig:LC_ORBS_8s} (see caption for details).

After the second peak discussed above, the flux decreases by a factor of $\sim$5 in 2~min when the observations where interrupted to observe the calibrator at around 05:35 UTC. As mentioned above, there are no detections neither before nor after the flare except for the three datapoints at around 05:37:30 UTC that appear to show marginal detections with a signal-to-noise ratio between 6$<$S/N$<$7. 
This is the most extreme event in the sample showing variability by a factor $\geq$10 followed by the sources 86 and 87 discussed later in section \ref{sec:variablity_all}, however these two additional sources show such variability on the opposite extreme on timescales of hours to days.

Based on the flare rise time up to the first peak and its corresponding light travel time, we can estimate an upper limit for the size scale of the emitting region to have a radius r$<$0.5~au. Following this constraint, the intensity at the first peak would be equivalent to a brightness temperature of 0.5 MK \citep{gue02}, which in turn represents a lower limit and thus an additional indication for the presence of high-energy particles, nonthermal emission thus being a possibility for the detected radio emission. Since the observation were carried out in dual-polarization mode it is not possible to recover Stokes V information. We imaged Stokes Q intensity maps, instead, but no signal is detected above the rms noise levels at the position of ORBS. The limited information on Stokes parameters does not allow for a conclusive assessment of neither linear nor circular polarization for the flare emission.

The position of this flare object coincides with a previously reported radio flaring source (within $\sim$0.09~arcsec) referred to as ORBS (Orion radio burst source) detected at cm-wavelengths ($\lambda=1.3$~cm; $\nu=22.3$~GHz) with the VLA in $K$-band \citep{for08}. During these observations (July 1991) this source showed an order of magnitude increase in its peak flux density in a few hours with a maximum at 47~mJy~beam$^{-1}$ with this spectacular event marking the source's first radio detection. This study reports a double radio source at 8.4~GHz (VLA $X$-band) toward this position of which the closest one to the ORBS source (within $\sim$0.11~arcsec) seems to be the south-west component of this double radio source (source SW in Table 2 of \citealt{for08}) while the other component is coincident with the position of another millimeter source in our catalogue within 0.12~arcsec, source 75, which does not show clear signs of variability at any time resolution (VF$\leq$2). This double radio source, of which the  ORBS is the south-west component, had first been detected at 8.4~GHz with the VLA from observations conducted in April 1994 and described in \citet{men95}, see their Figure 4.

While ORBS was detected during a strong X-ray flare later on (COUP 647, \citealt{get05a}), there are no additional counterparts at neither optical nor infrared wavelengths suggesting that this is a still deeply embedded source. The position of the COUP counterpart is coincident with the position of the flare peak within $\sim$0.17~arcsec and is indicated with a green circle in Figure \ref{fig:orbs}. It has been also detected more recently at cm-wavelength with the VLA and VLBA within $\sim$0.02~arcsec and $\sim$0.04~arcsec, respectively \citep{for16, vargas2021, dzib2021}. Identified as source 180 in \citet{vargas2021} and source 198 in \citet{for16}, without significant variability (VF$\lesssim$2). The VLBA observations consisted of 4 epochs, yet this source was only detected in two of them underlining its highly variable nature along with its non-thermal component. Its apparent large VLBA proper motion suggests this source alone (only the south-west component of the 8.4~GHz double radio source mentioned above) may be a binary system with angular separation of $\sim$4~au at the distance of the ONC. Interestingly, among the few millimeter YSO flares in literature, such as V773 Tau A and DQ Tau, are also multiple systems whose flaring mechanism is thought to be caused by interbinary magnetospheric interaction \citep{massi2006,salter2008}. Moreover, two highly variable sources discussed here (ORBS and source 86) and also reported as nonthermal radio sources in the VLBA observations discussed in \citet{forbrich2021} are already 50~per~cent of the potential binaries in that VLBA sample \citep{dzib2021}. The upper limit angular size scale for the emitting region derived earlier is $\sim$1.3~mas and thus comparable to the beam size for the unresolved VLBA detection for this source.

Among the 3-mm sources detected by \citet{friedel2011} using CARMA, source C5 is coincident with the position of source ORBS within 0.2~arcsec. It is reported with a peak flux density of $4.73\pm0.76$~mJy~beam$^{-1}$ similar to its values around the peak of the flare found here. Remarkably, the peak flux reported in \citet{friedel2011} comprises several hours of integration between different tracks during 2010, which implies a prolonged bright peak emission (or persisting flares) during those observations. For this flux measurement they used the CARMA A-configuration which resulted in a synthesized beam size of 0.4$\times$0.35~arcsec$^2$ similar to the typical beam size in our observations. However while we did not resolve ORBS they report it as marginally resolved with a deconvolved source size of 0.4$\times$0.29~arcsec$^2$.

Source ORBS provides a remarkable example of how radio emission (and X-ray emission as well) during flare events in protostars and YSOs can be significantly more luminous than that of main sequence stellar flares. We can compare the millimeter ORBS flare to the flares of Proxima Cen at 1.3~mm also observed with ALMA \citep{macgregor2018, macgregor2021}. These were remarkable short-duration $\leq$1~min flares of orders of magnitudes change in peak flux density, representing an analogous flare to those studied here but from a more evolved source (M dwarf with spectral type M5.5V). While the bright Proxima Cen flares peaked at around $\sim$100~mJy (the two observed flare peaks in 2017 and 2019), these peaks would have not been detected in our ALMA observations of the ONC, and at $\sim$400~pc these peaks would be roughly equivalent to a $\sim$1~$\mu$Jy (with a central frequency at 1.3~mm). On the contrary, the ORBS millimeter flare of $\sim$5.5~mJy~beam$^{-1}$ would be as bright as $\sim$500~Jy~beam$^{-1}$ at the distance of Proxima Cen (1.3~pc). This translates into absolute radio luminosities of $\sim2\times10^{14}$ and $\sim1\times10^{18}$~ergs~s$^{-1}$~Hz$^{-1}$ for Proxima Cen and ORBS flares, respectively. These differences highlight the importance of a continued systematic search for such events in YSOs to better constrain the nature of the radio emission during flares this way provide a significant sample for modelling studies such as those of T-Tauri magnetospheres to model both radio and X-ray emission during flares \citep{wat19}. Additionally, this finding is also providing a caveat for the study of disk masses, where a case such as ORBS with an averaged peak flux density in the concatenated data completely dominated by a flare would lead to a completely incorrect disk mass estimate. For instance, the ORBS flare would translate into significant change in dust mass since the latter is proportional to the measured flux, $M_{dust}\propto S_{\nu}$, assuming a fixed dust temperature in disks (see equation 1 in \citealt{eisner2018}), and therefore the variability factor of this flare would also mean an order of magnitude difference in the estimated dust mass. While a continued monitoring of millimeter variability will improve the statistical sample size necessary to assess the impact of such variability on disk mass estimates, our observations alone can already demonstrate that within a cumulative observing time of $\sim1276$~h (8 epochs of $\sim1.2$~h each with 133 sources) we find evidence of at least three out of 133 sources whose millimeter variability could lead to incorrect disk mass estimates with up to an order of magnitude difference in the estimated dust mass.


\subsubsection{Variability distribution in Orion-KL}\label{sec:variablity_all}

\begin{figure*}
    \centering
	\includegraphics[width=\linewidth]{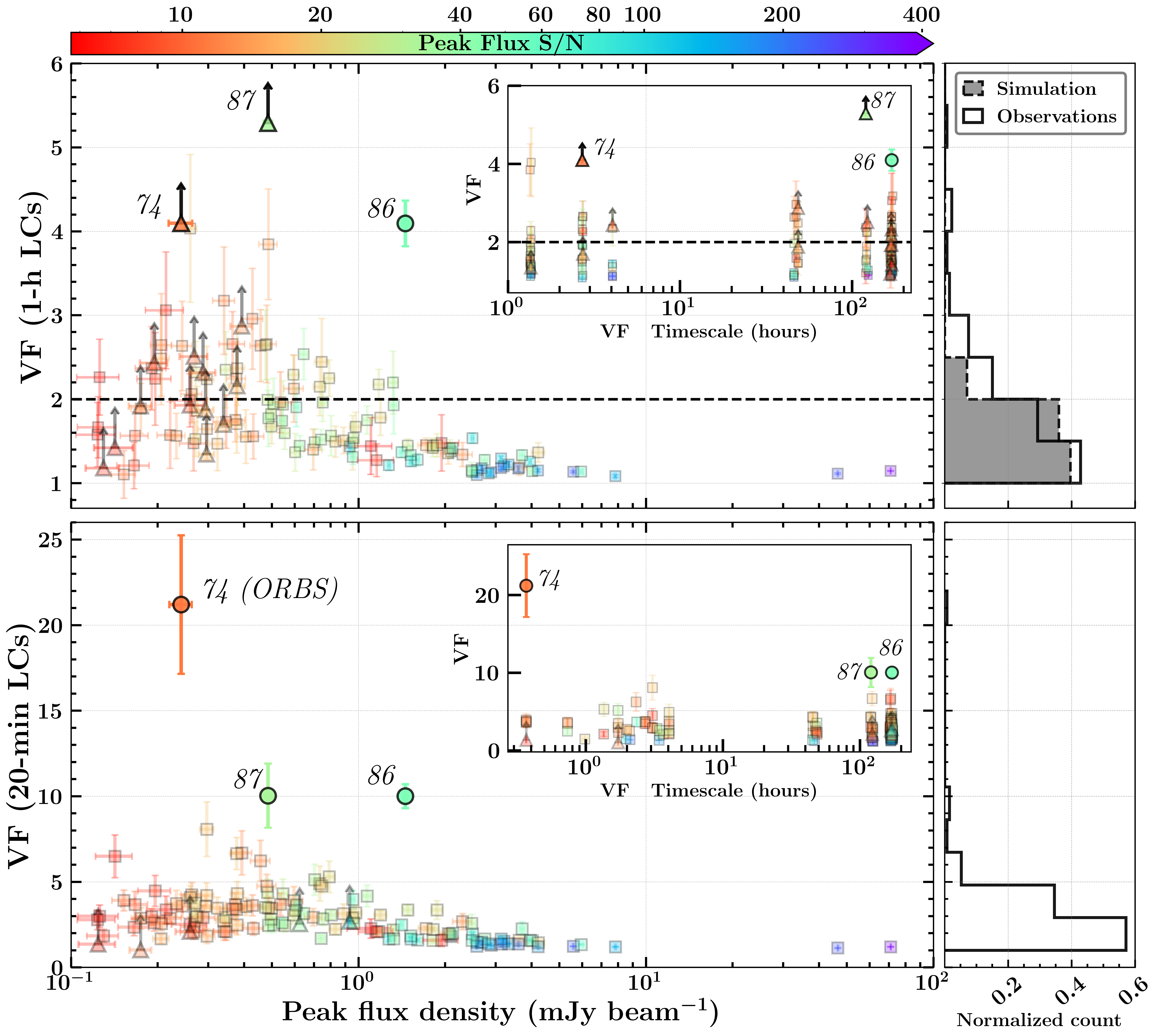}
    \caption{VF distribution at 1-h (top) and 20-min (bottom) time resolution  for the full sample as a function of peak flux density from the concatenated data with 1$\sigma$ error bars in both axis and color-coded by signal-to-noise ratio of the peak from the concatenated data. Lower limits are indicated by triangles and black arrows. The horizontal dashed line in the top panel indicates the systematic variability cut-off at VF = 2 described in Section \ref{sec:alma_sim}. Sources with VF above this cut-off within 3$\sigma$ uncertainty and also above the noise distribution shape (see text) are highlighted, others are shown with small square symbols and higher transparency levels. The full distribution histograms for both observations and simulations (described in section \ref{sec:alma_sim}) are shown on the right. The insets indicate the VF as a function of timescale for such variability level (time interval between the maximum and minimum in the LCs).}
    \label{fig:VF_dist}
\end{figure*}

As previously seen, already between the individual 1-h epochs alone we find variability occurring at all the analysed timescales, and even stronger events are accessible at shorter timescales. The resulting VF distributions from the LCs at 1-h and 20-min time resolution are shown in Figure \ref{fig:VF_dist} in the top and bottom panels, respectively, as a function of the averaged peak flux density from the concatenated data. While there is a wide range of variability at the two different time resolutions, the VF distribution from the LCs at shorter timescales reveals significantly greater variability levels of up to a factor of VF$\sim20$. This widespread variability occurs at all range of timescales (see insets in Figure \ref{fig:VF_dist}) with the strongest flare displayed by source 74 (ORBS) arising in less than an hour (bottom panel in Figure \ref{fig:VF_dist}). The other two sources with VF$\sim$10 (from the LCs at 20-min time resolution) show such variability on longer timescales of hours to days where a more prolonged flaring state may last longer than the observations (as seen in Figure \ref{fig:LCs}). 

These differences in the VF distribution when determining variability at different time resolutions are testimony to the interplay between the characteristic timescales of the variability in the sample and the averaged time intervals used to investigate this. For instance, a short-lived flare would be more evident if imaged or analyzed at a time resolution comparable or shorter than the duration of the event but then its signal would be progressively diluted within the average as longer time intervals are used to determine its brightness. We can see, for example, how short-lived substructures are seen in the LC of source 86 at higher time resolution (red area in the right-hand panel in Figure \ref{fig:LCs}) where the lower time resolution LC in the left-hand panel does not show clear evidence of the peak in sub-epoch 7 neither of the sudden increase in sub-epoch 12. A more obvious example was discussed for the ORBS and seen in Figures \ref{fig:LC_ORBS} and \ref{fig:LC_ORBS_8s} with LCs at different time resolutions showing the strong flare whose true maximum peak flux density significantly increases when measured at the shortest time resolutions and its detection in the concatenated data is only due to its strong short-lived flare.

Given the definition of the VF that is describing relative variability, greater variability levels are displayed predominantly towards lower averaged peak flux densities as seen in Figure \ref{fig:VF_dist}. Sources with averaged peak flux density above $\sim$2.0~mJy~beam$^{-1}$ are essentially constant and are most likely dominated by dust millimeter emission. Source BN, for example, is a well known thermal radio source \citep{for08,for16} and it is thus expected to show no signs of variability on short timescales. This is the brightest millimeter source in our sample (source 25 in Table \ref{tab:catalog}) with an averaged peak flux density of $70.991\pm0.183$~mJy~beam$^{-1}$ and shows indeed no millimeter variability with VF$\sim$1.1 and VF$\sim$1.2 from the 1-h and 20-min LCs, below our cut-off for potential systematic effects. While 98~per~cent of the sources in our catalogue have averaged peak flux density $<$8~mJy~beam$^{-1}$ the only other source with considerably bright peak flux density is the well known Source I (source 61 in our catalog) which together with source BN are the most massive objects in the Kleinmann–Low (KL) nebula in Orion within a range of $\sim$8$-$15~$M_\odot$ \citep{ginsburg2018, bally2020, wright2022}. Here we report an averaged peak flux density of $46.475\pm0.190$~mJy~beam$^{-1}$ for Source I and it is also amongst the most constant sources with VF$\sim$1.1 in both 1-h and 20-min time resolutions LCs.

If we only consider the VF$=2$ threshold discussed in section \ref{sec:alma_sim}, then, in the 1-h time resolution data, of the 133 LCs only $\sim$6~per~cent of the sources have VF values above our systematic limit of VF$>$2 within 3$\sigma$ uncertainty, including those with only upper limits available.  
As seen in the histograms on the top right-hand panel in Figure \ref{fig:VF_dist}, most of the sources show VF values below the defined cutoff at 1-h time resolution. The overall VF from the simulations is also shown in grey-filled histogram as reference. On the other hand, at 20-min time resolution, about $\sim$20~per~cent of the sample, present VF levels above this defined threshold within 3$\sigma$ uncertainty, including those with only upper limits available.

Furthermore, the envelope of the VF distribution at the two different time resolutions shown in Figure \ref{fig:VF_dist} also appears to display a systematic effect where a large dispersion is still seen above VF$=2$ , which is described by a noise distribution that increases towards lower peak flux densities and peaks at around $S_{\nu}\sim0.3-0.4$~mJy~beam$^{-1}$, reaching VF$\sim3-4$ at 1-h time resolution and VF$\sim5-7$ at 20-min time resolution. This envelope then decreases towards the faintest sources likely due to completeness issues and a selection bias where the faintest sources were mostly selected by visual inspection of the concatenated image, and these are located in regions less affected by background noise and therefore less likely to display large flux variations caused by noise. The envelope is indicated with transparent symbols in Figure \ref{fig:VF_dist}).  This effect appears to be driven by the VF definition, since the noise in a given LC has a larger impact on the measured VF for fainter sources. For example, for a constant source, a small flux fluctuation due to the dispersion in the LC for fainter $S_{\nu}$ values leads to higher VF compared to the impact of such dispersion in the LC of a constant bright source, leading to a systematically widening VF envelope towards lower flux densities in Figure \ref{fig:VF_dist}.

However, in our quest to quantify strong relative variability, the main goal is to identify sources that are undoubtedly variable beyond any systematic effect. This approach is partly motivated by the epoch-to-epoch ALMA flux calibration accuracy, which is also defined in relative terms (see for example \citealp{francis2020} and references therein). We thus follow the same approach as for the VF$=2$ threshold discussed in section \ref{sec:alma_sim}, where sources below this level, or within the dispersion envelope in this context, can still be variable, even though they are within the noise in the distribution. The variability would need to be checked individually, since it will already depend on whether there is complex image structure in the immediate vicinity. Even with these constraints we still find several sources that are clearly variable above any systematic effect: sources 74 (ORBS), 86, and 87. These three sources are above all systematic effects at both time resolutions, and the latter two are discussed individually in the following section, while source ORBS was already discussed in section \ref{sec:orbs}.

An additional note associated with the sample of disks that have been characterized at radio wavelengths with 3-mm flux measurements available in \citet{otter2021}, we can compare their flux measurements for those within the HPBW primary beam (47 sources). 57~per~cent of them have flux measurements compatible within 3$\sigma$ uncertainty, and 19~per~cent (9 sources) are not just incompatible within 3$\sigma$ uncertainty but also show at least 50~per~cent difference in flux density where the largest difference is shown by source 86 by a factor of $\sim$3.6 brighter in our observations (using the flux measurement from the concatenated data). This source is among the most variable ones in our sample, and is discussed in the following section together with source 87.


\subsubsection{Additional highly variable sources}\label{sec:individual_sources}

\begin{figure*}
    \centering
	\includegraphics[width=\linewidth]{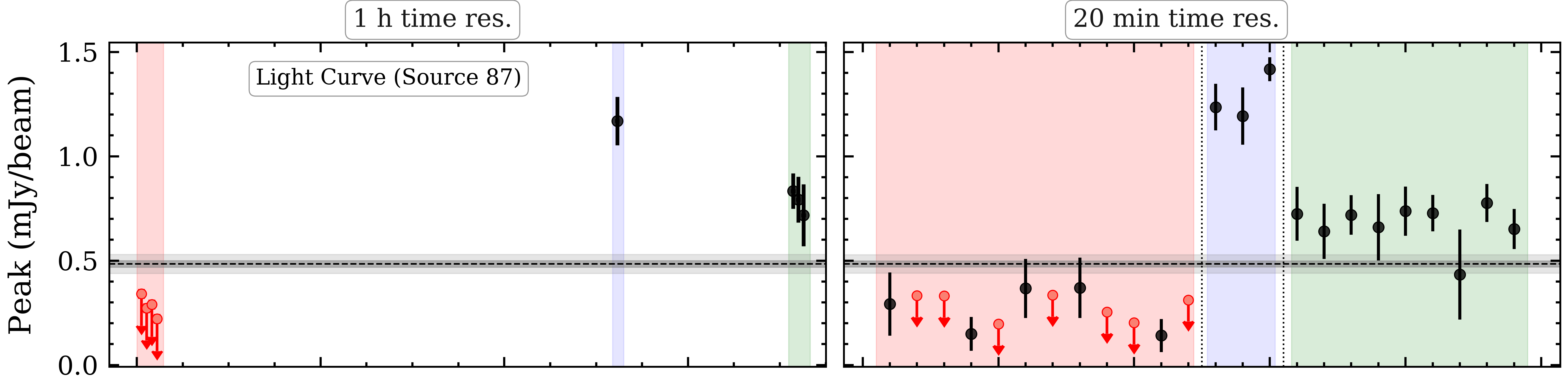}
	\includegraphics[width=\linewidth]{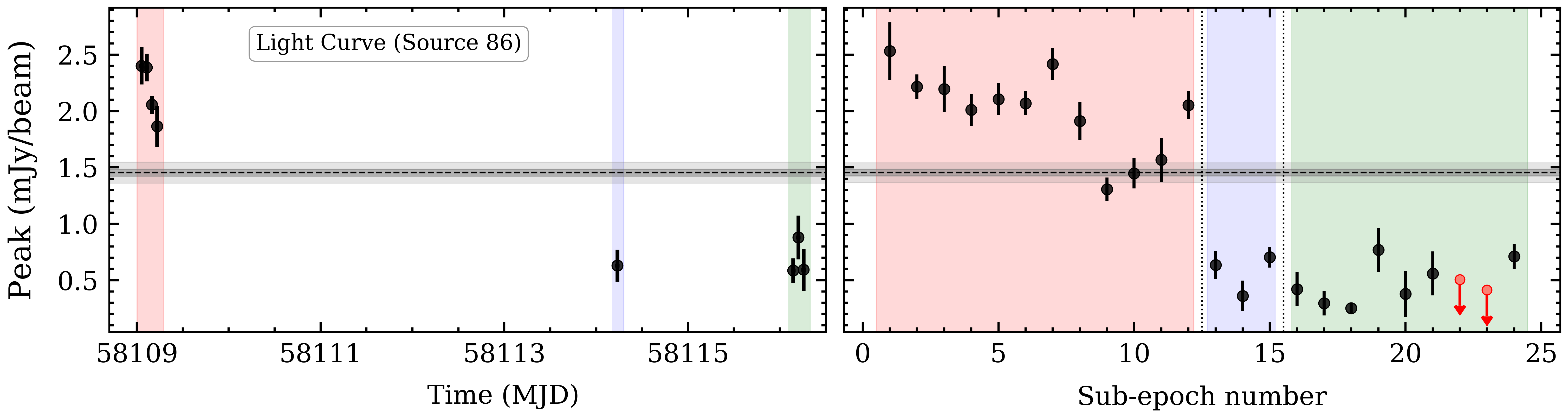}
    \caption{Radio LCs of the variable sources 86 and 87 at 1-h (left) and 20-min (right) time resolutions. Symbols and colors follows same notation as in Figure \ref{fig:LC_ORBS}.}
    \label{fig:LCs}
\end{figure*}

Here we will briefly comment on two additional sources showing the highest variability in the sample after the strong flare earlier discussed. These are the sources 86 and 87 whose largest variability occur on the longer timescales of days with changes in peak flux density by an order of magnitude or greater.

{\it \bf Source 86}: With a VF~$=4.1\pm0.3$ within 7~days (170.2~h) from its 1-h time resolution LC shown in the bottom left-hand panel in Figure \ref{fig:LCs}, this source shows a decreasing brightness in the first hours of observation (first 4 epochs, red are in Figure \ref{fig:LCs}) with a maximum peak flux density of $2.401\pm0.055$~mJy~beam$^{-1}$ in epoch 1 and a minimum in epoch 6 (first measurement within green area) at $0.586\pm0.037$~mJy~beam$^{-1}$. Interestingly, at shorter timescales, its 20-min time resolution LC illustrates the interplay between the averaged interval and the characteristic timescale of the corresponding variability, where an evident substructure begins to be temporally resolved revealing three successive peaks at $2.531\pm0.085$, $2.418\pm0.047$, and $2.052\pm0.042$~mJy~beam$^{-1}$, respectively, with the second and third peak rising after 2.1 and 2.7~h from the preceding peak where the maximum change in peak flux density for that interval (red area) occurs between the first and ninth measurement with a VF$\sim$1.9$\pm$0.1.

This source shows an order of magnitude change in peak flux density (VF$\sim$10.0$\pm$0.7) on longer timescales of 7.1~days (171~h) with a maximum at the beginning of the observations (sub-epoch 1) and a minimum within epoch 6 (sub-epoch 18, green band in Figure \ref{fig:LCs}). From sub-epoch 13 onward the LC fluctuates around $0.463$~mJy~beam$^{-1}$ with a standard deviation of $0.166$~mJy~beam$^{-1}$. Similarly, this source was also detected at 3~mm continuum observations from September 2017 \citep{otter2021, ginsburg2018} with a reported flux density of $0.409\pm0.004$~mJy from aperture photometry (see Table 5 in \citealt{otter2021}, source 37), which in line with our measurements from sub-epoch 13 onward, may represent a quiescent state of source 86, with the caveat that even such ``quiescent state'' may still be dominated by flares. An example of this can be illustrated by the seemingly ``quiescent state'' of source 87 in its 20-min resolution LC starting from sub-epoch 16 onward (top right-hand panel in Figure \ref{fig:LCs}, green area) where its peak flux density displays a ``quiescent'' constant level (except for sub-epoch 22) yet brighter than the first twelve sub-epochs (red area) which are otherwise upper limits mostly. 

Source 86 was reviously reported at cm-wavelengths as GMR D in \citealt{gar87}, and more recently in \citealt{for16}, and \citealt{vargas2021} with no significant variability (sources [FRM2016] 211 and [VFD2021] 186, respectively). It was reported as a nonthermal radio source in \citealt{forbrich2021} where its VLBA unusual proper motion suggests these are detections of different components among the observations \citep{dzib2021} and in such a case source 86 would actually be a close binary system. It also has an X-ray counterpart (COUP 662) with a hydrogen column density log$(N_{H})=23.22\pm0.03$ leading to a high visual extinction $A_V\sim80$ (using the conversion $N_H/A_V=2\times10^{21}$~cm$^{-2}$ from \citealt{vuong2003}), which supports the fact that neither optical nor IR counterparts have been reported for this source.

{\it \bf Source 87}: This source shows the largest variability in the sample at 1~h time resolution with a VF$\sim$5.3 on a timescale of 5~days (120.2~h) and it is indeed just a lower limit variability since its true radio luminosity remained below detectable levels during the first four epochs where the minimum is reported (as three times the rms noise; top left panel in Figure \ref{fig:LCs}). The local rms noise at the minimum (epoch 4) is 0.074~mJy~beam$^{-1}$. The averaged peak flux density of source 87 over the 8 epochs is 0.484$\pm$0.015~mJy~beam$^{-1}$, but peaks at $1.169\pm0.039$~mJy~beam$^{-1}$ in epoch 5 (blue band in Figure \ref{fig:LCs}), almost 2.5 times brighter than the averaged peak flux density. If we then look into the 20-min time resolution LC, its peak within epoch 5 increases to $1.417\pm0.019$~mJy~beam$^{-1}$ leading to a variability of an order of magnitude (VF$=10.0\pm1.9$) in 5~days.

This source has been previously detected at cm-wavelengths in \citet{for16} and \citet{vargas2021} (source [FRM2016] 212 and [VFD2021] 187, respectively). Source [VFD2021] 187 shows a decrease in peak flux density by a factor of $\sim3$ in nearly 2~h as measured from its 5-min time resolution LC of the central pointing presented in \citealt{vargas2021}. It has an X-ray counterpart in the COUP survey (COUP 670) with a reported spectral type between K4-M0 and a visual extinction $A_V\sim2.31$ (based on optical and infrared properties from \citealt{hillenbrand1997, luhman2000, lucas2001}) and a near-IR counterpart in the VISION survey (VISION 05351492-0522392; \citealt{mei16}). 
Also reported as a nonthermal radio source at cm-wavelengths with the VLBA, where it was only detected in one out of four observed epochs with a 35.1$\sigma$ significance level \citep{dzib2021,forbrich2021} pointing out to its extreme variability in the cm-range. Further evidence of its millimeter variability can be inferred from similar ALMA 3mm observations conducted three months prior to our observations where no peak above a 5$\sigma$ detection threshold is found on images with reported rms noise levels between 0.04 $-$ 1.0~mJy~beam$^{-1}$ \citep{otter2021}. According to its IR counterpart in \citealt{muench2002} (source 568, with an angular separation of $\sim$0.14~arsec), \citealt{otter2021} determined a 3mm upper limit a this position of 0.027~mJy (three times the local rms noise).


\section{Summary and Conclusions}\label{sec:conclusions}

We present ALMA 3mm continuum observations towards the Orion BN/KL region at sub-arcsecond resolution and report the first systematic search for mm-wavelengths flares in YSOs on timescales from minutes to days. 

We detect 133 sources within a area of $\sim$1.6$\times$1.6~arcsec$^2$ ($\sim$0.2$\times$0.2~pc$^2$) and have studied their LCs at different time resolutions. Within this sample, we report the discovery of a strong flare from a known YSOs previously reported as a radio flaring source detected at cm-wavelengths and referred to as ORBS where it showed an order of magnitude change in peak flux density in just a few hours \citep{for08}.  In our ALMA observations it was only detected in one of the 8 epoch (individual epochs of 1~h each). This single detection and only at this time resolution corresponds to a change in peak flux density by a factor of at least $>$4 in less than three hours. Further analysis of this flare at high-time resolution of 8-s cadence allowed us to constrain the development of this strong event that had a duration of $\sim$10~min with more than an order of magnitude change in peak flux density in $\sim$4~min. At this high-time resolution we are also able to resolve a lightcurve substructure at the peak of the event where a double peak is seen at $\sim$4.9 and $\sim$5.8~mJy~beam$^{-1}$ separated by 3~min. This strong millimeter flare from a known YSO is a remarkable evidence of how radio time domain analysis of such dataset is providing us with a new perspective on high-energy irradation of YSO vicinities, its impact on protoplanetary disks and ultimately on planet formation.

Radio variability analysis for a dataset of this kind towards a complex region such as the ONC necessarily requires time-slicing imaging for flux measurements at different time resolutions. This method entails some uncertainties for lower levels of variability where systematic effects are expected mostly due to the impact that a dynamic shape and size of the synthesized beam throughout the observations has on flux measurements of both resolved sources and/or unresolved sources in a complex region with a variable background. Using simulated observations, we conclude that these systematic effects could produce artificial variability of up to a factor of VF$\sim$2. 

Finally, this study is providing a first look at the capabilities that ALMA offers to the field of radio time-domain studies at high-time resolution in the millimeter range, which also has an impact on the interpretation of averaged millimeter fluxes, such as in the study of disk masses for individual YSOs. Additionally, our findings provide strong evidence of the value of both continued radio monitoring of YSOs and the development of even more efficient methods for the analysis of variability in such radio datasets of complex regions at high-time resolution which undoubtedly provides a unique window to the study of high-energy processes at the earliest phases of stellar evolution.

\section*{Acknowledgements}

We thank the anonymous referee for a constructive review of this paper. This research made use of Astropy,\footnote{\url{http://www.astropy.org}} a community-developed core Python package for Astronomy \citep{astropy13, astropy18} and Matplotlib \citep{hun07}; APLpy, an open-source plotting package for Python (\citealp{aplpy2012, robitaille_thomas_2019_2567476} - \url{https://github.com/aplpy/aplpy}); The University of Hertfordshire high-performance computing facility (\url{http://stri-cluster.herts.ac.uk}). V.M.R. has received support from the Comunidad de Madrid through the Atracción de Talento Investigador Modalidad 1 (Doctores con experiencia) Grant (COOL:Cosmic Origins of Life; 2019-T1/TIC-5379), and the Ayuda RYC2020-029387-I funded by MCIN/AEI /10.13039/501100011033.

\section{Data Availability}

This paper makes use of the following ALMA data:\#2017.1.01313.S. ALMA is a partnership of ESO (representing its member states), NSF (USA), and NINS (Japan), together with NRC (Canada), NSC and ASIAA (Taiwan), and KASI (Republic of Korea), in cooperation with the Republic of Chile. The Joint ALMA Observatory is operated by ESO, AUI/NRAO, and NAOJ.

\bibliographystyle{mnras}
\bibliography{orion} 

\appendix

\bsp	
\label{lastpage}
\end{document}